\documentclass[english,prl,twocolumn]{revtex4-1}
\usepackage[T1]{fontenc}
\usepackage[latin9]{inputenc}
\usepackage{amsmath}
\usepackage{amssymb}
\usepackage{graphicx}
\usepackage{babel}
\begin{document}
\title{Anomalous Coherence Length of Majorana Zero Modes at Vortices in Superconducting
Topological Insulators}
\author{Bo Fu and Shun-Qing Shen}
\affiliation{Department of Physics, The University of Hong Kong, Pokfulam Road, Hong Kong, China}
\date{\today}
\begin{abstract}
The coherence length of two Majorana zero energy modes in a p-wave topological superconductor
is inversely proportional to the superconducting order parameter. We studied the
finite size effect of the Majorana zero modes at vortices in a topological insulator/superconductor
heterostructure in the presence of a vortex and found that the the coherence length
of the two zero energy modes at the terminals of a vortex line is independent of
superconducting order parameter, and determined by the intrinsic properties of the
topological insulator. This anomalous property illustrates that the superconducting
topological insulator is topologically distinct, contrary to a $p$-wave topological
superconductor.
\end{abstract}
\maketitle

\paragraph*{Introduction}

Search for Majorana zero modes in topological phases has generated extensive interest
in condensed matter physics and material science \citep{Wilczek-09np,Stern-10nature,Qi-11rmp,Alicea-12rpp,Shen-12book,Beenakker-13arcmp,Sato-16jpsj,Lutchyn-18nrm}.
The Majorana zero mode in topological superconductor carry zero energy and obey non-Abelian
statistics. Their occupancy can form the topological degeneracy of the ground states
of the system, which are expected to have potential application for fault tolerant
topological quantum computation \citep{Kitaev-03ap,Freedman-03,Nayak-08rmp}. In
their pioneering work, Fu and Kane \citep{Fu-08prl} proposed that the proximity
effect between an s-wave superconductor and the surface electrons of a strong topological
insulator leads to a time-invariant superconducting state resembling a spinless $p_{x}\pm ip_{y}$
superconductor. They proposed that this interface supports Majorana bound modes at
vortices. Over the last decade, this proposal has attracted significant attention
and become one of the main prototypes to construct and to engineer a physical system
to host the topological excitations \citep{Lutchyn-10prl,Oreg-10prl,Sau2010prl,Sau10prb,Mourik-15science,Das-12np,Nadj-Perge-14science,XuJP-15prl,Sun-16prl,Hu-16prb},
and to understand the zero energy modes observed in iron-based superconductors \citep{Zhang_-18Science,WangDF-18Science,Machida-19nm,KongLY,LiuQ-18prx,LiuWY-20nc,ZhuSY-20Science,Chiu-20sa,LiM-22nature,Konig-19prl,zhang-21prl,Kheirkhah-21prb,Hu-arxiv}.
However the search for Majorana zero modes is meeting a great of difficulty and challenge
especially in experiments.

The time-reversal-invariant superconductor with spin-orbit coupling belongs to symmetry
class DIII and in two-dimensions is characterized by a $\mathbb{Z}_{2}$ topological
invariant\citep{Schnyder-08prb,Kitaev-09aip,Chiu-16rmp}. The topologically nontrivial
phase hosts a pair of helical Majorana modes on its edge \citep{Fu2010odd,Qi-10prb,Zhang-13prl,Haim-19PhysRep}.
Theoretically, it can be realized by considering the spin-triplet superconducting
pairing with odd-parity or extended-s wave pairing which flips its sign when it evolves
across the Brillouin zone. Based on the odd-parity superconductivity criterion \citep{Fu2010odd,Sato2010Topological},
the strong topological insulator in contact with an s-wave superconductor is topologically
trivial without edge modes, which is distinctly different from the chiral $p$-wave
topological superconductor \citep{Green-00prb,Ivanov-01prl}. The existence of the
zero energy vortex mode in this system is associated with the Atiyah-Singer index
theorem which clarifies the correspondence between the vorticity of the vortex and
the number of the localized zero-energy mode for the surface states \citep{Atiyah-63bam,Jackiw-81npb,Jackiw-07prl}.
It heavily relies on the validity of the topological insulator's surface states and
the presence of the chiral symmetry. Furthermore, the chemical potential enters into
the Bogoliubov-de Gennes (BdG) equation in a nontrivial way and breaks the chiral
symmetry explicitly that the index theorem does not apply here. There arises the
question how the tunneling between two surfaces lifts the degeneracy of the Majorana
modes in a thin film for a finite chemical potential.

In the present work, we investigate the finite size effect of the Majorana zero energy
modes in vortices in the topological insulator/superconductor (TI/SC) heterostructure
depicted as Fig. 1(a). In the presence of the superconducting vortex the two Majorana
modes are present and connected through the bulk topological insulator along the
vortex when the chemical potential $\mu$ is lower than a critical value $\mu<\mu_{c}$.
The energy splitting of the two modes decays exponentially with the thickness and
the coherence length only depends on the intrinsic properties of the topological
insulator, and is independent of the superconducting order parameter. As a comparison,
we also present the results for the semimagnetic topological insulator/ superconductor
(SMTI/SC) heterostructure depicted as Fig. 2(a), which is equivalent to a $p$-wave
topological superconductor when the chemical potential locates within the magnetic
gap of the top surface states. The two Majorana modes reside at the vortex core and
at the boundary separately, and their coherence length is equal to the superconducting
coherence length, which is a typical signature of a p-wave topological superconductor.
Thus the anomalous coherence length in the TI/SC heterostructure indicates that the
the pair of the zero energy modes at the vortex core is attributed to the winding
number of the superconducting order parameter, not to the p-wave topological superconductivity.

\begin{figure*}
\includegraphics[width=16cm]{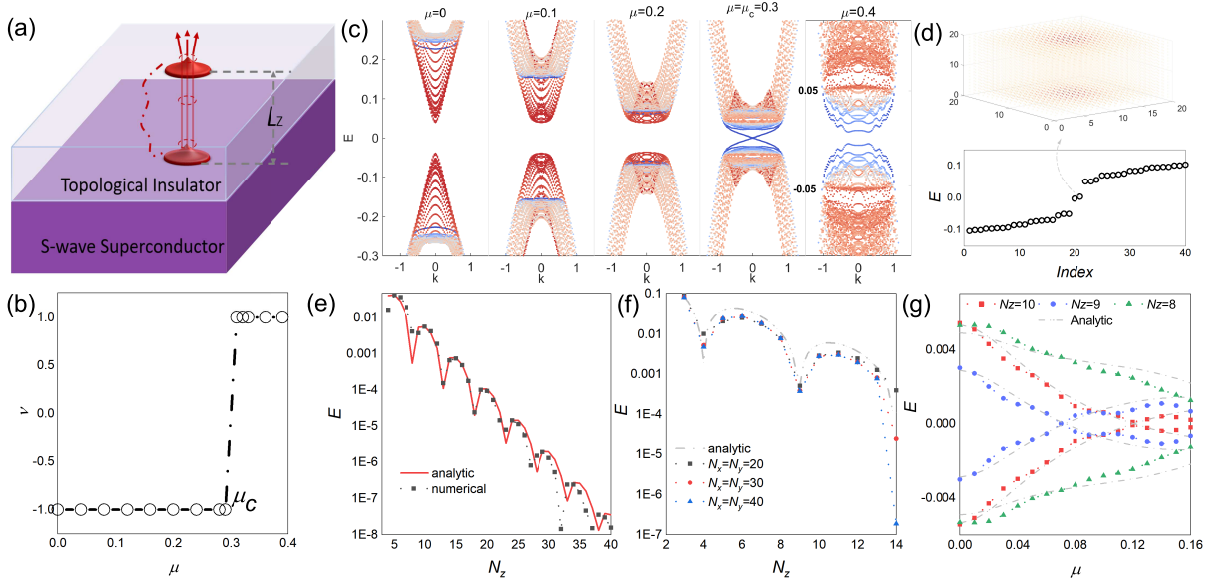} \caption{The TI/SC heterostructure in the presence of a superconducting vortex. (a) Schematic
with two zero energy modes bound to a vortex line. $L_{z}$ denotes the thickness
of the sample. (b) $\mathbb{Z}_{2}$ topological invariant as a function of $\mu$.
(c) The dispersions of quasi-1D system at different values of $\mu$. The blue-to-red
color gradient indicates the radial probability distribution for each band with the
vortex line at the origin. The blue colored bands correspond the dispersions for
the vortex bound states. (d) Energy spectrum and the wave-function for the zero energy
modes for open boundary conditions. (e) Plot of the finite size induced gap between
the surface states at $k_{x}=k_{y}=0$ as a function of the thickness $L_{z}=N_{z}a$
with $a$ as the lattice constant. (f) For $\mu=0$, the evolution of the energy
of the vortex bound states with respect to the thickness $N_{z}$ (g) The the energies
of the vortex states as a function of $\mu$. For $N_{z}=8$, $N_{x}=N_{y}=50$ and
for $N_{z}=9,10$, $N_{x}=N_{y}=40$. Parameters are $\Delta=0.05$, $\hbar v=0.4$,
$B=0.5$, and $m=0.28$.}
\end{figure*}

\paragraph*{The TI/SC heterostructure}

We start with a minimal bulk model for a three dimensional topological insulator
$H_{\mathbf{k}}$ which supports gapless surface states \citep{Qi-11rmp}. The numerical
calculation is based on a tight-binding model on a cubic lattice, and the analytical
study is based on the continuum model in the long wavelength approximation that the
lattice Hamiltonian is expanded in terms of the wave vector $\mathbf{k}$ to the
second order around $\Gamma$ point, $H_{\mathbf{k}}=v\rho_{x}\mathbf{k}\cdot\boldsymbol{\sigma}+M(\mathbf{k})\rho_{z}-\mu$
where $M(\mathbf{k})=m-B\mathbf{k}^{2}$, the Pauli matrices $\sigma$ and $\rho$
are acting on spin and orbit space respectively\citep{ZhangH-09np,Qi-11rmp,Shen-12book},
$v,m,B$ are the material parameters and $\mu$ is the chemical potential. In proximity
to an $s$-wave superconductor, a finite superconducting pairing $\Delta$ is induced
in the three-dimensional topological insulator which leads to the BdG Hamiltonian
$H_{\mathbf{k}}^{\mathrm{BdG}}$. Based on the odd-parity superconductivity criterion,
$H_{\mathbf{k}}^{\mathrm{BdG}}$ is topologically trivial without gapless edge modes
around the system \citep{Fu2010odd}. In the presence of a vortex in $z$ direction,
$\Delta\to\Delta(r)e^{i\theta}$ with $(r,\theta)$ as the in-plane polar coordinates
with respect to the vortex core, the translational invariance along $z$ direction
persists and $k_{z}$ is still a good quantum number. The problem becomes to classifying
the gapped phases in quasi-1D whose unit cell consists all the sites in $xy$ plane.
Due to the lacking of time reversal symmetry, it belongs to symmetry class D and
is characterized by a $\mathbb{Z}_{2}$ invariant $\nu$ which is defined as the
product of signs of Pfaffians of the antisymmetric and real BdG Hamiltonians in the
Majorana representation at two time reversal invariant momenta $k_{z}=0,\pi$ \citep{Kitaev-09aip,Wimmer-12ACM}.
With increasing $\mu$, the quasi-1D system transitions into the trivial phase via
a quantum critical point $\mu=\mu_{c}\simeq v\sqrt{\frac{m}{B}}$ at which $\nu$
changes from $-1$ to $+1$ as shown in Fig. 1(b). In Fig. 1(c), we also present
the numerical results of the evolution of the dispersion under the variation of $\mu$.
The results reveal that the vortex line is fully gapped except at $\mu=\mu_{c}$
where a vortex phase transition takes place\citep{Hosur-11prl}. For $\nu=-1$ ($\mu<\mu_{c}$),
the quasi-1D system is topological nontrivial and there exists a single 0D zero mode
at each end of the termination along $z$ direction\citep{Hosur-11prl,Chiu-11prb}
as shown in Fig. 1(d). In order to estimate the energy splitting of two vortex line
end states for finite thickness $L_{z}$, we derive the effective Hamiltonian for
the vortex line. By projecting onto the two states centered at the vortex line which
is denoted by the darkest of blue color in Fig. 1(c), we obtain the effective dispersions
(See Appendix B)
\begin{equation}
H_{\mathrm{eff}}=-\mathcal{F}(k_{F}^{2}\xi^{2})\left[(\widetilde{m}-Bk_{z}^{2})\nu_{z}+vk_{z}\nu_{y}\right]\label{eq:eff_H}
\end{equation}
where $\widetilde{m}=m-Bk_{F}^{2}-\frac{B}{\xi^{2}}$ is the renormalized mass with
$k_{F}=\mu/v$ and $\xi=v/\Delta$, $\nu_{i}$ are Pauli matrices acting on the projected
two bands and $\mathcal{F}(x)=[-\frac{\mathrm{E}(-x)}{x+1}+\mathrm{K}(-x)]/[\mathrm{E}(-x)-\mathrm{K}(-x)]$
is monotonically decreasing function where $\mathrm{K}$ and $\mathrm{E}$ are the
complete elliptic integral of the first and second kind, respectively. The superconducting
pairing only enters into an overall energy renormalization function $\mathcal{F}$
which can be factored out without changing any of the topological properties. Then,
the effective Hamiltonian (\ref{eq:eff_H}) resembles the 1D Su--Schrieffer--Heeger
model \citep{Su-80prb} instead of the 1D topological Kitaev chain in which that
the p-wave pairing is linear in momentum \citep{Kitaev06ap}. Consequently, when
$\widetilde{m}B>0$, the vortex line Hamiltonian is topologically nontrivial. At
$\widetilde{m}=0$ which corresponds a critical chemical $\mu=\mu_{c}$, the gap
vanishes signaling a topological phase transition which is consistent with the $\mathbb{Z}_{2}$
topological invariant. From Eq. (\ref{eq:eff_H}), we can obtain the energy splitting
for the zero energy modes
\begin{equation}
\delta E=\frac{4\widetilde{m}v\mathcal{F}(k_{F}^{2}\xi^{2})}{\sqrt{4B\widetilde{m}-v^{2}}}\left|\sin\left(\frac{\sqrt{4\widetilde{m}B-v^{2}}}{2B}L_{z}\right)\right|\exp\left(-\frac{vL_{z}}{2B}\right).\label{eq:fsz_vortex_line}
\end{equation}
For the chemical potential $\mu\sim0$ or in the strong pairing limit $k_{F}\xi\ll1$,
we have $\mathcal{F}(k_{F}^{2}\xi^{2})\simeq1$ and $\widetilde{m}\simeq m$, the
energy splitting is independent on the superconducting pairing and recovers the finite
size effect for the surface state of topological insulator \citep{Zhou-08prl,Linder-09prb}.
As shown in Fig. 1(f), the energy splitting of the zero energy states based on a
tight-binding numerical calculations quickly saturates when the size of the slab
is much large than the superconducting coherence length $L_{x},L_{y}\gg\xi$ and
features an oscillating exponential decay with increasing the thickness of the sample,
which agrees well with the analytic expression. For $\mu\ne0$, since the chemical
potential enters into oscillating function $\sin\left(\frac{\sqrt{4\widetilde{m}B-v^{2}}}{2B}L_{z}\right)$
through the renormalized mass $\widetilde{m}$, the energy splitting is also sensitive
to $\mu$ besides thickness of the sample as shown in Fig. 1(g). For large $\mu$
or the weak pairing limit $k_{F}\xi\gg1$, $\mathcal{F}(k_{F}^{2}\xi^{2})\simeq\frac{\ln(4k_{F}\xi)-1}{k_{F}^{2}\xi^{2}}$
shows a power-law decrease of $k_{F}\xi$. Due to the presence of this pre-factor,
the finite size effect is strongly suppressed.

This energy splitting can be understood from aspect of top and bottom surface states
with superconducting pairing by means of the index theorem \citep{Weinberg-81prd,Fukui-10jpsj,Roy-14prb}.
The Dirac surface states of strong topological insulator thin films can be described
by $h_{\mathbf{k}_{\shortparallel}}^{\mathrm{surf}}=v\varrho_{z}(\mathbf{k}_{\shortparallel}\times\boldsymbol{\sigma})_{z}+t\varrho_{x}-\mu$
where $\mathbf{k}_{\shortparallel}=(k_{x},k_{y},0)$ denotes the in-plane wave vector,
$t$ is the inter-surface tunneling and the Pauli matrices $\varrho$ and $\sigma$
denote surface and spin degrees of freedom, respectively.. In combination with the
superconducting pairing, the BdG Hamiltonian is $H_{\mathrm{BdG}}^{\mathrm{surf}}(\mathbf{k}_{\shortparallel})=\tau_{z}h_{\mathbf{k}_{\shortparallel}}^{\mathrm{surf}}+\Delta\tau_{x}$
which belongs to the Altland-Zirnbauer symmetry class DIII and is classified by a
$\mathbb{Z}_{2}$ topological invariant. This system has the mirror symmetry $M_{z}=i\varrho_{x}\sigma_{z}$
with $M_{z}^{2}=-1$ which reflects the top surface to the bottom surface. After
a uniform $\pi/2$ rotation around $\varrho_{y}\sigma_{z}$, the full BdG Hamiltonian
can be decoupled into the direct sum of two mirror sectors $H_{\mathbf{k}_{\shortparallel}}^{\chi}$
with mirror eigenvalue as $i\chi$. The two subblocks are particle-hole partners
of each other $\tau_{y}\sigma_{x}H_{\mathbf{k}_{\shortparallel}}^{\chi*}\sigma_{x}\tau_{y}=-H_{-\mathbf{k}_{\shortparallel}}^{-\chi}$.
Each subblock breaks particle-hole symmetry explicitly and possesses the chiral symmetry
$\{\mathcal{C},H_{\mathbf{k}_{\shortparallel}}^{\chi}\}=0$ with $\mathcal{C}=\tau_{y}$
thus belongs to the class AIII. In two spatial dimensions, the topological classification
for class AIII is trivial \citep{Schnyder-08prb}. In the presence of a vortex, the
pertinent Hamiltonian for $\mu=0$ becomes
\begin{equation}
H^{\chi}=\chi\tau_{z}\left[-iv\left(\sigma_{y}\partial_{x}-\sigma_{x}\partial_{y}\right)+t\sigma_{z}\right]+\Delta\left(\cos\theta\tau_{x}+\sin\theta\tau_{y}\right)
\end{equation}
where $\mathbf{k}_{\shortparallel}$ is replaced by $-i(\partial_{x},\partial_{y})$.
Note that the five four-dimensional Hermitian matrices anticommute with each other
and the interface tunneling term enters into the Hamiltonian as the fifth anti-commuting
matrix. For $t=0$, there is additional chiral symmetry $\tau_{z}\sigma_{z}$ in
$H^{\chi}$ which can be expressed as $\{\tau_{z}\sigma_{z},H^{\chi}\}=0$ and ensures
the spectral symmetry. As a consequence, the zero energy states $|\Psi_{0}^{\chi}\rangle$
of $H^{\chi}$ becomes eigenstates of $\tau_{z}\sigma_{z}$ with eigenvalue as $+1$.
The analytic index of the chiral symmetric model is defined by $\mathrm{ind}H^{\chi}=n_{+}-n_{-}$where
$n_{\pm}$ are the number of zero-energy states with chirality $\pm1$. The index
theorem states that the analytic index is identical to the winding number of the
the order parameter in the two-dimensional space and there are exactly $n$ number
of zero modes for the vorticity $n$ \citep{Weinberg-81prd,Fukui-10jpsj,Roy-14prb}.
Also as pointed in Ref. \citep{Teo-10prl}, the zero modes are associated with hedgehogs
in the complex vector fields of the superconducting order parameter $\mathbf{n}(\mathbf{r})=(\Delta\cos\theta,-\Delta\sin\theta)$.
In particular, when the vorticity is one there exist single state at zero energy
for $H^{\chi}$. After including the chirality symmetry breaking term $\chi t\tau_{z}\sigma_{z}$,
we have $(H^{\chi}+\chi t\tau_{z}\sigma_{z})|\Psi_{0}^{\chi}\rangle=\chi t\tau_{z}\sigma_{z}|\Psi_{0}^{\chi}\rangle=\chi t|\Psi_{0}^{\chi}\rangle$
that the energy will be shifted from zero to $\chi t$. The existence of the zero
energy solution heavily relies on the assumption that the intersurface tunneling
is negligible. When the tunneling effects are taken into account, the zero energy
bound states are actually shifted away from zero.

\begin{figure*}
\includegraphics[width=16cm]{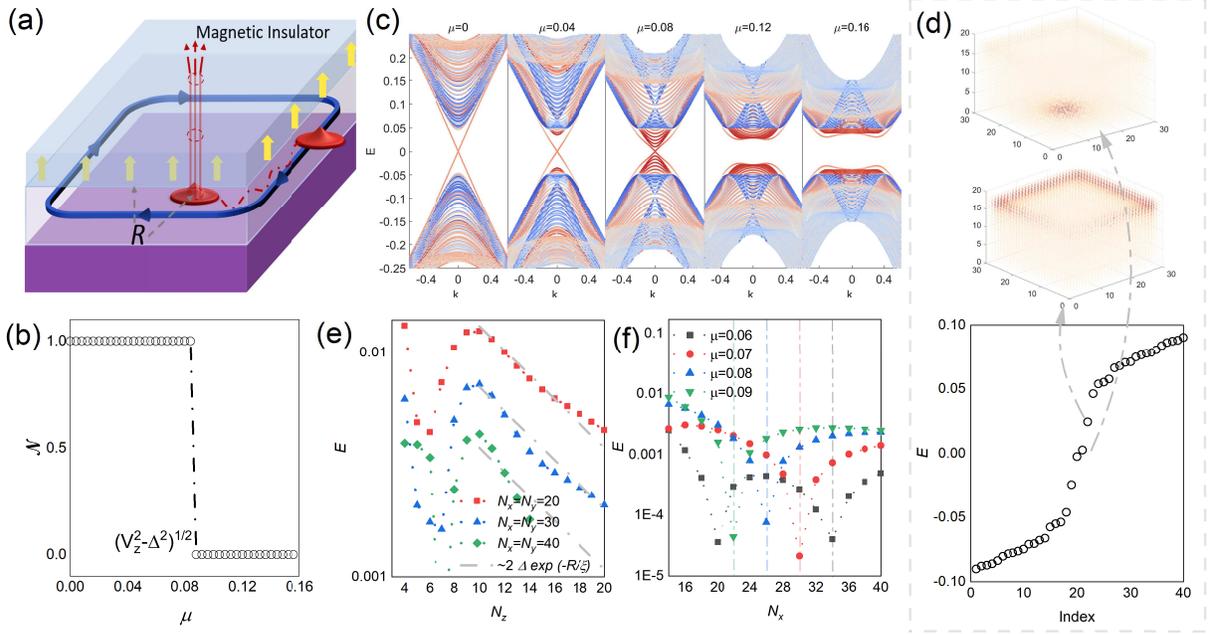}\caption{The SMTI/SC heterostructure. (a) Schematic of the zero energy modes residing in the
vortex core and the boundary. (b) In the absence of vortex, the Chern number $\mathcal{N}$
of the quasi-2D system as a function of $\mu$. (c) The quasi-one-dimensional band
structure for $\mu$ for $N_{z}=10$ and $N_{y}=100$. The distribution of the wave
function along the $z$ direction indicated by the scale from blue to yellow color.
(d) The energy spectra for the open boundary condition and the illustration of the
spatial probability distribution $|\Psi(x,y,z)|^{2}$ of zero energy state and the
chiral state for a $30\times30\times18$ lattice. (e) At $\mu=0$, the energy for
the lowest energy mode as a function of $N_{z}$ for different $N_{x}$ and $N_{y}$.
(f) For finite $\mu$, the energy for the lowest energy mode as a function of $N_{x}$
and $N_{y}$ for $N_{z}=10$. The singularity points in the logarithmic plot due
to the oscillation are indicated by the vertical dashed lines. Parameters are $\Delta=0.05$,
$\hbar v=0.4$, $b=0.5$, and $m=0.28$. The exchange field $V_{z}=0.1$ are added
to top three layers.}
\end{figure*}

\paragraph*{The SMTI/SC heterostructure}

For comparison, we now turn to the SMTI/SC heterostructure. The exchange interaction
between the magnetic ion and the surface electrons leads to nonzero magnetization
and makes the top surface electrons open an energy gap $2|V_{z}|$. When the Fermi
level intersects gapless bottom surface states and locates within the magnetic gap
of the top surface states, i. e. $|\mu|<\sqrt{V_{z}^{2}-\Delta^{2}}$, the quasi-2D
system is topologically equivalent to a chiral topological superconductor with nonzero
Chern number supporting chiral Majorana modes on its boundary \citep{Fu-08prl,Akhmerov-09prl,Fu-09prl,qi-10prb-b,Teo-10prb,Qi-13prb-c,Wang-15prb,wang-16prb,Lian-18PNAS,Yan2021majorana}.
A topological phase transition occurs at $|\mu|=\sqrt{V_{z}^{2}-\Delta^{2}}$ in
Fig. 2(b). It is verified by numerical calculations for different $\mu$ as shown
in Fig. 2(c). The presence of edge states is consistent with the bulk band topology.
After introducing a vortex, the existence of the Majorana zero energy state is governed
by the BdG Hamiltonian for the surface states. We then map the surface states onto
a 2D plane with the intersection point of the vortex line with the bottom surface
mapped to the center of the plane. In this situation, the exchange field only exists
outside a disk radius $R$, i. e. $M(r)=V_{z}\Theta(r-R)$. For $\sqrt{V_{z}^{2}-\Delta^{2}}>\mu>0$,
we find a zero energy solution $|\psi_{\mathrm{core}}\rangle$ localized at the vortex
core and a bound state $|\psi_{\mathrm{inter}}\rangle$ localized at the interface
between the magnetic and nonmagnetic regions \citep{Ivanov-01prl,Green-00prb,Law-09prl}.
We can construct the approximate eigenstate wave function as $|\Psi_{\pm}\rangle=\frac{1}{\sqrt{2}}\left(|\psi_{\mathrm{core}}\rangle\pm|\psi_{\mathrm{inter}}\rangle\right)$
with the energies $E_{+}=-E_{-}=\delta E$ \citep{Cheng-09prl,Cheng-10prb}. These
two wave functions satisfy the particle-hole symmetry of the BdG equations, $\Xi|\Psi_{+}\rangle=|\Psi_{-}\rangle$.
As shown in Fig. 2(d), based on the tight binding calculations, we present the energy
spectra and the wave functions for the lowest two energy states. The zero energy
state is constituted by two parts of contributions: one part is exponentially localized
at the vortex core while the other part is localized at the interface. In addition
to the zero modes, there are chiral modes peaked only at the interface within the
superconducting gap. By considering the overlapping of two zero energy modes, the
energy splitting can be obtained as (Appendix C)
\begin{equation}
\delta E\approx2\Delta e^{-R/\xi}|\sin(k_{F}R-\delta)|\label{eq:fsz_for_zeeman}
\end{equation}
where $\delta=\mathrm{arctan}\sqrt{\frac{V_{z}+\mu}{V_{z}-\mu}}+\frac{\pi}{4}$.
The energy splitting decays exponentially as a function of $R$. The coherence length
is simply the superconducting coherence length $\xi=v/\Delta$, which is proportional
inversely to the superconducting order parameter $\Delta$. The analytic result (\ref{eq:fsz_for_zeeman})
is in qualitative agreement with the numerical results as shown in Fig. 2(e) and
(f).

In order to gain better insight into the difference between the two present cases,
we derive an effective model to capture the main physics in the SMTI/SC heterostructure.
We start from the effective Hamiltonian for the strong topological insulator thin
films in contact with the magnetic insulator on its top, $h_{\mathbf{k}_{\shortparallel}}^{\mathrm{mag}}=-v(\mathbf{k}_{\shortparallel}\times\boldsymbol{\sigma})_{z}+M(\mathbf{k}_{\shortparallel})\sigma_{z}$
\citep{Mogi-21np,Zou-22-arxiv}. At low energy ($k_{\shortparallel}<k_{c}$ with
$k_{c}\simeq\sqrt{\frac{m}{B}}$), $M(\mathbf{k}_{\shortparallel})=0$, $h_{\mathbf{k}_{\shortparallel}}^{\mathrm{mag}}$
turns out to be the massless Dirac Hamiltonian which describes the bottom surface
states. At high energy regime ($k_{\shortparallel}>k_{c}$), $M(\mathbf{k}_{\shortparallel})\ne0$
originates from the surface states evolve into the bulkin the high energy regime
and breaks time reversal symmetry explicitly. Thus, $M(\mathbf{k}_{\shortparallel})$
behaves as a regularization term and change the band topology. The wave function
with the positive eigenvalue can be solved as $|\psi_{c}\rangle=\left(i\cos\frac{\varphi_{\mathbf{k}_{\shortparallel}}}{2},e^{i\theta_{\mathbf{k}_{\shortparallel}}}\sin\frac{\varphi_{\mathbf{k}_{\shortparallel}}}{2}\right)$
for $M(\infty)>0$ and $|\psi_{c}\rangle=\left(ie^{-i\theta_{\mathbf{k}_{\shortparallel}}}\cos\frac{\varphi_{\mathbf{k}_{\shortparallel}}}{2},\sin\frac{\varphi_{\mathbf{k}_{\shortparallel}}}{2}\right)$
for $M(\infty)<0$ where $\cos\varphi_{\mathbf{k}_{\shortparallel}}=M(\mathbf{k}_{\shortparallel})/\epsilon\text{(\ensuremath{\mathbf{k}_{\shortparallel}})}$
with $\epsilon\text{(\ensuremath{\mathbf{k}_{\shortparallel}})}=\sqrt{v^{2}k_{\shortparallel}^{2}+M^{2}(\mathbf{k}_{\shortparallel})}$.
Note that the angular factor $e^{\pm i\theta}$ must accompany with the component
vanishing at $k_{\shortparallel}\to\infty$ to ensure the single-valueness of the
wave function. After the inclusion of the superconducting pairing, the BdG Hamiltonian
can be projected onto the two bands which are intersected with the Fermi energy,
\begin{equation}
H_{\mathrm{\mathrm{BdG}}}^{\mathrm{mag}}=\left(\begin{array}{cc}
\epsilon\text{(\ensuremath{\mathbf{k}_{\shortparallel}})}-\mu & \frac{vk_{\shortparallel}}{\epsilon\text{(\ensuremath{\mathbf{k}_{\shortparallel}})}}\Delta e^{-\mathrm{sgn}[M(\infty)]i\theta_{\mathbf{k}_{\shortparallel}}}\\
\frac{vk_{\shortparallel}}{\epsilon\text{(\ensuremath{\mathbf{k}_{\shortparallel}})}}\Delta e^{\mathrm{sgn}[M(\infty)]i\theta_{\mathbf{k}_{\shortparallel}}} & -\epsilon\text{(\ensuremath{\mathbf{k}_{\shortparallel}})}+\mu
\end{array}\right).
\end{equation}
It is an effective chiral $p$-wave BdG Hamiltonian and the chirality crucially depends
on the sign of the $M(\infty)$. The projected Hamiltonian cannot be determined without
ambiguity in the absence of the regulator. The Chern number for the occupied band
is $\mathcal{N}=\mathrm{sgn}[M(\infty)]$. The presence of a vortex will leads to
the anti-periodical condition for the wave function, $\Psi(r,\theta+2\pi)=e^{i\pi}\Psi(r,\theta)$\citep{Green-00prb,Ivanov-01prl}.
By using the Ansatz for the wave function $\Psi_{l}(\mathbf{r})=\frac{e^{il\theta}}{\sqrt{2\pi r}}[e^{-i\theta/2}u_{l}(r),e^{i\theta/2}v_{l}(r)]$,
we can obtain the the radial BdG Hamiltonian for $[u_{0}(r),v_{0}(r)]$
\[
H_{\mathrm{BdG}}^{\mathrm{radial}}=[\epsilon(-\partial_{r}^{2})-\mu]\tau_{z}-i\frac{\Delta}{2\mu}v\partial_{r}\tau_{x}
\]
which is equivalent to 1D Kitaev chain. The superconducting pairing only enters into
the off-diagonal terms in sharp contrast with Eq. (\ref{eq:eff_H}).

\paragraph*{Summary}

The pairing patterns of the Majorana modes in the TI/SC and SMTI/SC heterostructure
in the presence of a vortex have different features. Both numerical simulation and
analytical analysis show that the energy splitting of the modes decays exponentially
in the thickness $L_{z}$ of the TI layer in the TI/SC structure and the size $R$
of the SMTI/SC structure. The coherence length in the TI/SC is independent of the
superconducting order parameter, which is contrary to that in the SMTI/SC structure
or a p-wave topological superconductor. The distinct behaviors of the coherence lengths
in two cases reveal that the microscopic origins and the topological nature of the
vortex Majorana zero modes are different.
\begin{acknowledgments}
This work was supported by the National Key R\&D Program of China under Grant No.
2019YFA0308603 and the Research Grants Council, University Grants Committee, Hong
Kong under Grant Nos. C7012-21GF and 17301220.
\end{acknowledgments}

\section*{Appendix}

\subsection*{A. The tight-binding model for numerical simulations}

In this section, we give the explicit model for the tight-binding calculations. The
microscopic model for the bulk model of the hybrid system with a single vortex is
defined by the Hamiltonian,

\[
\hat{H}_{\mathrm{tot}}=\hat{H}_{\mathrm{TI}}+\hat{H}_{\mathrm{SC}}+\hat{H}_{\mathrm{Z}}.
\]
The $\hat{H}_{\mathrm{TI}}$ term describes the topological electronic structure
of the bulk system. In the basis of $|\mathrm{P}1_{z}^{+},\uparrow\rangle$,$|\mathrm{P}1_{z}^{+},\downarrow\rangle$
$|\mathrm{P}2_{z}^{-},\uparrow\rangle$, $|\mathrm{P}2_{z}^{-},\downarrow\rangle$,
it can be written as \citep{ZhangH-09np,Shen-12book,Qi-11rmp}
\begin{align*}
\hat{H}_{\mathrm{TI}} & =\sum_{\mathbf{r}}\psi_{\mathbf{r}}^{\dagger}\left[\left(M-\sum_{\boldsymbol{\delta}}B_{\boldsymbol{\delta}}\right)\rho_{z}\sigma_{0}-\mu\rho_{0}\sigma_{0}\right]\psi_{\mathbf{r}}\\
 & +\sum_{\mathbf{r},\boldsymbol{\delta}}\left[\psi_{\mathbf{r}}^{\dagger}\frac{1}{2}\left(B_{\boldsymbol{\delta}}\rho_{z}\sigma_{0}-\frac{iv_{\boldsymbol{\delta}}}{a}\rho_{x}\sigma_{\delta}\right)\psi_{\mathbf{r}+\boldsymbol{\delta}}+h.c.\right]
\end{align*}
where $\rho_{i}$ and $\sigma_{i}$ ($i=0,x,y,z$) are Pauli matrices act on the
orbit and spin spaces, respectively. $\psi_{\mathbf{r}}=[c_{\mathbf{r}1\uparrow},c_{\mathbf{r}1\downarrow},c_{\mathbf{r}2\uparrow},c_{\mathbf{r}2\downarrow}]^{T}$
are annihilation operators of the four-component spinor at position $\mathbf{r}$.
$2M$ is the band gap at $\Gamma$ point. $\frac{\hbar}{a}v_{\boldsymbol{\delta}}$
and $B_{\boldsymbol{\delta}}$ describes the spin-dependent and spin-independent
hoppings on the cubic lattice along $\boldsymbol{\delta}$ direction with $\boldsymbol{\delta}=x,y,z$.
$a$ is the lattice constant. $\mu$ is the chemical potential. For simplicity, we
take $B_{x}=B_{y}=B_{z}=B$ then the above model describes a strong topological insulator
phase with gapless surface states for $2>M/B>0$.

The proximity-induced superconductivity $\hat{H}_{\mathrm{SC}}$ can be described
by,
\[
\hat{H}_{\mathrm{SC}}=\sum_{\mathbf{r}}\left[\psi_{\mathbf{r}}^{\dagger}\Delta(\mathbf{r})e^{-i\phi(\mathbf{r}_{\parallel})}i\sigma_{y}\psi_{\mathbf{r}}^{\dagger}+h.c.\right]
\]
 where $\phi(\mathbf{r}_{\parallel})=\mathrm{Arg}(\mathbf{r}_{\parallel}-\mathbf{r}_{\parallel}^{0})$
with $\mathbf{r}_{\parallel}=(x,y)$ as the planar position vector, $\mathbf{r}_{\parallel}^{0}$
as the coordinate for the vortex line and $\mathrm{Arg}$ representing the the argument
of the vector. The pairing function is written as $\Delta(\mathbf{r})=\Delta_{0}f(z)\tanh(r_{\parallel}/\xi)$
where $\Delta_{0}$ is the amplitude of the pairing function, $f(z)$ and $\tanh(r_{\parallel}/\xi)$
are the distribution function along $z$ and the planar direction respectively, and
$\xi=v/\Delta_{0}$ is the coherent length of the superconductor. Here we think the
thickness of topological insulator film is much less than the coherent length of
the superconductor such that $f(z)\simeq1$.

$\hat{H}_{Z}$ describes the Zeeman term which is modeled as 
\[
\hat{H}_{\mathrm{Z}}=\sum_{\mathbf{r}}V_{z}(z)\psi_{\mathbf{r}}^{\dagger}\rho_{0}\sigma_{z}\psi_{\mathbf{r}}
\]
with $V_{z}$ as the amplitude of the exchange field. The exchange field is only
restricted to several layers near the top surface of the system, $V_{z}(z)=V_{z}\Theta(z_{m}-z)$
where $\Theta$ is Heaviside step function and $z_{m}$ is the thickness of magnetic
layers.

\subsection*{B. Derivation of the effective vortex Hamiltonian}

The inclusion of the s-wave superconductivity in three dimensional topological insulator
leads to the Bogoliubov--de Gennes (BdG) Hamiltonian as

\[
H_{\mathbf{k}}^{\mathrm{BdG}}=\left(\begin{array}{cc}
H_{\mathbf{k}} & \Delta\\
\Delta & -\sigma_{y}H_{-\mathbf{k}}^{*}\sigma_{y}
\end{array}\right),
\]
where $\Delta$ is the superconducting pairing. In the presence of a vortex in $z$
direction, the superconducting pairing $\Delta\to\Delta(r)e^{i\theta}$ with respect
to the vortex core. We have used cylindrical coordinates $(x,y,z)=(r\cos\theta,r\sin\theta,z)$.
In view of the rotational symmetry about $z$ axis, we can assign the quantum numbers
$(k_{z},l,n)$ for the bulk of the system, which are the momentum in $z$-direction,
the angular momentum in $xy$ plane and the radial quantum number respectively. These
emergent symmetries of the effective Hamiltonian allow us to obtain the radial BdG
equation $H_{k_{z},l}^{\mathrm{BdG}}\Psi_{k_{z},l,n}=E_{k_{z},l,n}\Psi_{k_{z},l,n}$
at a given $k_{z}$ and $l$ with 
\[
H_{k_{z},l}^{\mathrm{BdG}}=\left(\begin{array}{cc}
H_{k_{z},l} & \Delta(r)\\
\Delta(r) & -\sigma_{y}H_{-k_{z},-l}^{*}\sigma_{y}
\end{array}\right)
\]
where
\[
H_{k_{z},l}=\left(\begin{array}{cccc}
M_{k_{z}}^{1-l}-\mu & 0 & vk_{z} & -viD_{r}^{l}\\
0 & M_{k_{z}}^{l}-\mu & -viD_{r}^{1-l} & -vk_{z}\\
vk_{z} & -viD_{r}^{l} & -M_{k_{z}}^{1-l}-\mu & 0\\
-viD_{r}^{1-l} & -vk_{z} & 0 & -M_{k_{z}}^{l}-\mu
\end{array}\right).
\]
Here we have introduced $D_{r}^{l}=\partial_{r}+l/r$ and $M_{k_{z}}^{l}=m-Bk_{z}^{2}+BD_{r}^{1-l}D_{r}^{l}$.
To find the effective Hamiltonian for the vortex line, we first rewrite the Hamiltonian
as $H(k_{z},l)=H_{1}(\partial_{r},l)+H_{2}(k_{z},l)$, then solve solutions for $k_{z}$-independent
part $H_{1}$and finally project $k_{z}$-dependent part $H_{2}$ onto the relevant
bands. $H_{1}$ can be expressed as a direct sum of $+$ and $-$ sectors $H_{1}^{\pm}(\partial_{r},l)=-iv(\partial_{r}+\frac{1}{2r})\tau_{z}\nu_{x}-\mu\tau_{z}+\Delta(r)\tau_{x}\mp\frac{v}{2r}\tau_{0}\nu_{y}\pm\frac{vl}{r}\tau_{z}\nu_{y}$.
For $l=0$, there exists a chiral symmetry $\tau_{y}\nu_{x}H_{1}^{\pm}(\partial_{r},l=0)\tau_{y}\nu_{x}=-H_{1}^{\pm}(\partial_{r},l=0)$
which plays an essential role in the determination of the zero energy solutions.
The zero energy solutions for $H_{1}^{\pm}$ can be solved as
\begin{align*}
|\phi_{+}\rangle & =Ne^{-\int_{0}^{r}dr^{\prime}\frac{\Delta(r^{\prime})}{v}}\left(\begin{array}{c}
-J_{1}(k_{F}r)\\
0\\
0\\
iJ_{0}(k_{F}r)\\
J_{0}(k_{F}r)\\
0\\
0\\
iJ_{1}(k_{F}r)
\end{array}\right),\\
|\phi_{-}\rangle & =Ne^{-\int_{0}^{r}dr^{\prime}\frac{\Delta(r^{\prime})}{v}}\left(\begin{array}{c}
0\\
J_{0}(k_{F}r)\\
iJ_{1}(k_{F}r)\\
0\\
0\\
J_{1}(k_{F}r)\\
-iJ_{0}(k_{F}r)\\
0
\end{array}\right)
\end{align*}
with the normalization factor defined as 
\[
4\pi N^{2}\int_{0}^{\infty}drre^{-2\int_{0}^{r}dr^{\prime}\frac{\Delta(r^{\prime})}{v}}[J_{0}^{2}(k_{F}r)+J_{1}^{2}(k_{F}r)]=1.
\]
The expectation values of the remaining terms in $H_{2}$ with respect to the zero
energy solutions for $H_{1}^{\pm}$ can be calculated as
\begin{align*}
\langle\phi_{s}|\mathcal{M}_{k_{z}}^{0}|\phi_{s^{\prime}}\rangle & =s\delta_{ss^{\prime}}\mathcal{F}(k_{F}^{2}\xi^{2})\left[m-B(k_{z}^{2}+k_{F}^{2}+\frac{1}{\xi^{2}})\right],\\
\langle\phi_{s}|vk_{z}\tau_{z}\rho_{x}\sigma_{z}|\phi_{s^{\prime}}\rangle & =-s\delta_{s,-s^{\prime}}ivk_{z}\mathcal{F}(k_{F}^{2}\xi^{2})
\end{align*}
where $s,s^{\prime}=\pm$ and $\mathcal{M}_{k_{z}}^{0}=\rho_{z}\otimes\mathrm{diag}(M_{k_{z}}^{1},M_{k_{z}}^{0},-M_{k_{z}}^{0},-M_{k_{z}}^{1})$,
which leads to Eq. (1) in main text.

\subsection*{C. Zero energy solutions for SMTI/SC heterostructure}

In this section, we calculate the zero energy solutions for SMTI/SC heterostructure
and the energy splitting due to their overlapping. In polar coordinates, by using
the Ansatz for the wavefunction $\Psi_{l,n}(r,\theta)=e^{i[l-\frac{1}{2}(\tau_{z}+\sigma_{z})]\theta}\Psi_{l,n}(r)$
with $l$ being an integer to make it single-valuedness, we can separate angular
and radial variables. The radial Hamiltonian is rewritten as $H_{l}(r)\Psi_{l,n}(r)=E_{l,n}\Psi_{l,n}(r)$
with 
\begin{equation}
H_{l}(r)=\left(\begin{array}{cc}
h_{l}(r) & \Delta(r)\\
\Delta^{*}(r) & -\sigma_{y}h_{-l}^{*}(r)\sigma_{y}
\end{array}\right)
\end{equation}
where $h_{l}(r)$ is given by 
\begin{equation}
h_{l}(r)=\left(\begin{array}{cc}
-\mu+M(r) & -ivD_{r}^{l}\\
-ivD_{r}^{1-l} & -\mu-M(r)
\end{array}\right).
\end{equation}
Due to the presence of the particle-hole symmetry $H_{l}=-\Xi H_{-l}\Xi^{-1}$with
$\Xi=\tau_{y}\sigma_{y}\mathcal{K}$, if $E$ is an eigenvalue with the eigenfunction
$\Psi_{l,E}(r)=[u_{\uparrow},u_{\downarrow},v_{\downarrow},v_{\uparrow}]^{T}$, then
$-E$ is also an eigenvalue and the corresponding eigenfunction is $\Psi_{-l,-E}(r)=[-v_{\uparrow}^{*},v_{\downarrow}^{*},u_{\downarrow}^{*},-u_{\uparrow}^{*}]^{T}$.
The zero energy state only exists for $l=0$ and needs to be an eigenstate of $\Xi$,
i.e. $\Xi\Psi_{0,0}(r)=\zeta\Psi_{0,0}(r)$, which gives constraints on components
of the eigenfunction $u_{\uparrow}=-\zeta v_{\uparrow}^{*}$ and $u_{\downarrow}=\zeta v_{\downarrow}^{*}$.By
redefining the spinor $u_{\uparrow}=\widetilde{u}_{\uparrow}e^{-i\pi/4-\zeta\int_{0}^{r}dr^{\prime}\Delta(r^{\prime})/\hbar v}$
and $u_{\downarrow}=\widetilde{u}_{\downarrow}e^{i\pi/4-\zeta\int_{0}^{r}dr^{\prime}\Delta(r^{\prime})/\hbar v}$
, the four coupled differential equations in radial BdG equations is reduced to two
real equations 
\begin{align}
-[\mu-M(r)]\widetilde{u}_{\uparrow}+v\partial_{r}\widetilde{u}_{\downarrow} & =0,\nonumber \\
v\left(\partial_{r}+\frac{1}{r}\right)\widetilde{u}_{\uparrow}+[\mu+M(r)]\widetilde{u}_{\downarrow} & =0.\label{eq:zero_mode}
\end{align}
For $0<\mu<V_{z}$, by solution Eq. (\ref{eq:zero_mode}), we find a solution localized
at the vortex core, 
\[
\left(\begin{array}{c}
u_{\uparrow}\\
u_{\downarrow}
\end{array}\right)_{\mathrm{core}}\sim e^{-i\pi/4\sigma_{z}-\int_{0}^{r}dr^{\prime}\frac{\Delta(r^{\prime})}{\hbar v}}\left(\begin{array}{c}
J_{1}(k_{F}r)\\
-J_{0}(k_{F}r)
\end{array}\right),
\]
with the positive eigenvalue $\zeta=+1$ of $\Xi$ and a bound state localized at
the interface 
\begin{align}
\left(\begin{array}{c}
u_{\uparrow}\\
u_{\downarrow}
\end{array}\right)_{\mathrm{inter}} & \sim\frac{e^{-i\pi/4\sigma_{z}}}{\sqrt{r}}e^{\int_{R}^{r}dr^{\prime}\frac{\Delta(r^{\prime})}{\hbar v}}\nonumber \\
 & \times\begin{cases}
\left(\begin{array}{c}
\sqrt{\frac{\mu+V_{z}}{2V_{z}}}\\
\sqrt{\frac{V_{z}-\mu}{2V_{z}}}
\end{array}\right)e^{-\frac{\sqrt{m^{2}-\mu^{2}}(r-R)}{\hbar v}}, & r>R,\\
\left(\begin{array}{c}
-\sin\left[k_{F}(r-R)-\delta+\frac{\pi}{4}\right]\\
\cos\left[k_{F}(r-R)-\delta+\frac{\pi}{4}\right]
\end{array}\right), & r<R,
\end{cases}
\end{align}
with the negative eigenvalue $\zeta=-1$ of $\Xi$ where $\delta=\arctan\sqrt{\frac{V_{z}+\mu}{V_{z}-\mu}}+\frac{\pi}{4}$
is a phase determined by matching the wavefunction at the interface. To estimate
the energy splitting for Majorana modes, we multiply $\langle\psi_{\mathrm{core}}|$
to the BdG equation $H_{0}(r)|\Psi_{+}\rangle=E_{+}|\Psi_{+}\rangle$ which yields
$E_{+}=\frac{\langle\psi_{\mathrm{core}}|H_{0}(r)|\Psi^{+}\rangle}{\langle\psi_{\mathrm{core}}|\Psi^{+}\rangle}$.
Then using the relation $H_{0}(r)|\psi_{\mathrm{inter}}\rangle=-iv\tau_{z}\sigma_{x}|\psi_{\mathrm{inter}}\rangle\delta(r-R)$
and $H_{0}(r)|\psi_{\mathrm{core}}\rangle=0$, we arrive the expression for the energy
splitting for the zero mode in the main text {[}Eq. (4){]}.

\end{document}